\documentclass[onecolumn,amsmath,amssymb,11pt,superscriptaddress,nofootinbib]{revtex4-1}
\pdfoutput=1

\usepackage[english]{babel}
\usepackage{amssymb}
\usepackage{amsmath}
\usepackage{amsthm}
\usepackage[]{graphicx}
\usepackage{tensor}
\usepackage{color}
\usepackage{framed}
\usepackage{caption}
\usepackage{natbib}
\usepackage{dcolumn}
\usepackage{bm}
\usepackage{subcaption}
\usepackage{enumitem}

\usepackage{fancyhdr}
\usepackage[us,12hr]{datetime}

\usepackage{graphicx}
\usepackage{epstopdf}

\usepackage[symbol]{footmisc}

\usepackage{tikz}
\usetikzlibrary{decorations.pathmorphing, patterns,shapes}
\usetikzlibrary{calc}
\usetikzlibrary{intersections}
\usetikzlibrary{angles}
\usetikzlibrary{quotes}

\input epsf

\begin{document}

\allowdisplaybreaks
\begin{titlepage}

\title{A strategy for finding people infected with SARS-CoV-2:\\ 
optimizing pooled testing at low prevalence
}
\author{Leon Mutesa}
\affiliation{Centre for Human Genetics, College of Medicine and Health Sciences, University of Rwanda, Kigali, Rwanda}
\affiliation{Rwanda Joint Task Force COVID-19,  Rwanda Biomedical Centre, Ministry of Health, Kigali, Rwanda}
\author{Pacifique Ndishimye}
\affiliation{African Institute for Mathematical Sciences, Kigali, Rwanda}
\affiliation{Rwanda Joint Task Force COVID-19, Rwanda Biomedical Centre, Ministry of Health, Kigali, Rwanda}
\author{Yvan Butera}
\affiliation{Rwanda Joint Task Force COVID-19,  Rwanda Biomedical Centre, Ministry of Health, Kigali, Rwanda}
\author{Jacob Souopgui}
\affiliation{Centre for Human Genetics, College of Medicine and Health Sciences, University of Rwanda, Kigali, Rwanda}
\affiliation{Rwanda Joint Task Force COVID-19,  Rwanda Biomedical Centre, Ministry of Health, Kigali, Rwanda}
\affiliation{Department of Molecular Biology, Institute of Biology and Molecular Medicine, IBMM, Université Libre de Bruxelles, Gosselies, Belgium}
\author{Annette Uwineza}
\affiliation{Rwanda Joint Task Force COVID-19,  Rwanda Biomedical Centre, Ministry of Health, Kigali, Rwanda}
\author{Robert Rutayisire}
\affiliation{Rwanda Joint Task Force COVID-19,  Rwanda Biomedical Centre, Ministry of Health, Kigali, Rwanda}
\author{Emile Musoni}
\affiliation{Rwanda Joint Task Force COVID-19,  Rwanda Biomedical Centre, Ministry of Health, Kigali, Rwanda}
\author{Nadine Rujeni}
\affiliation{Rwanda Joint Task Force COVID-19,  Rwanda Biomedical Centre, Ministry of Health, Kigali, Rwanda}
\author{Thierry Nyatanyi} 
\affiliation{Rwanda Joint Task Force COVID-19,  Rwanda Biomedical Centre, Ministry of Health, Kigali, Rwanda}
\author{Edouard Ntagwabira}
\affiliation{Rwanda Joint Task Force COVID-19,  Rwanda Biomedical Centre, Ministry of Health, Kigali, Rwanda}
\author{Muhammed Semakula}
\affiliation{Rwanda Joint Task Force COVID-19,  Rwanda Biomedical Centre, Ministry of Health, Kigali, Rwanda}
\author{Clarisse Musanabaganwa}
\affiliation{Rwanda Joint Task Force COVID-19,  Rwanda Biomedical Centre, Ministry of Health, Kigali, Rwanda}
\author{Daniel Nyamwasa}
\affiliation{Rwanda Joint Task Force COVID-19,  Rwanda Biomedical Centre, Ministry of Health, Kigali, Rwanda}
\author{Maurice Ndashimye}
\affiliation{African Institute for Mathematical Sciences, Kigali, Rwanda}
\affiliation{Rwanda Joint Task Force COVID-19,  Rwanda Biomedical Centre, Ministry of Health, Kigali, Rwanda}
\author{Eva Ujeneza}
\affiliation{African Institute for Mathematical Sciences, Kigali, Rwanda}
\author{Ivan Emile Mwikarago} 
\affiliation{Rwanda Joint Task Force COVID-19,  Rwanda Biomedical Centre, Ministry of Health, Kigali, Rwanda}
\author{Claude Mambo Muvunyi} 
\affiliation{Rwanda Joint Task Force COVID-19,  Rwanda Biomedical Centre, Ministry of Health, Kigali, Rwanda}
\author{Jean Baptiste Mazarati} 
\affiliation{Rwanda Joint Task Force COVID-19,  Rwanda Biomedical Centre, Ministry of Health, Kigali, Rwanda}
\author{Sabin Nsanzimana}
\affiliation{Rwanda Joint Task Force COVID-19,  Rwanda Biomedical Centre, Ministry of Health, Kigali, Rwanda}
\author{Neil Turok}
\email{nturok@perimeterinstitute.ca}
\affiliation{African Institute for Mathematical Sciences, Kigali, Rwanda}
\affiliation{James Clerk Maxwell Building, University of Edinburgh, EH9 3FD, Edinburgh, Scotland}
\affiliation{Perimeter Institute for Theoretical Physics, 31 Caroline St N, Ontario, Canada}
\author{Wilfred Ndifon}
\email{wndifon@nexteinstein.org}
\affiliation{African Institute for Mathematical Sciences, Kigali, Rwanda}

\begin{abstract}
Suppressing SARS-CoV-2 will likely require the rapid identification and isolation of infected individuals, on an ongoing basis. RT-PCR (reverse transcription polymerase chain reaction) tests are accurate but costly, making regular testing of every individual expensive. The costs are a challenge for all countries and particularly for developing countries. Cost reductions can be achieved by pooling (or combining) subsamples and testing them in groups. We propose an algorithm for pooling subsamples based on the geometry of a hypercube that, at low prevalence, uniquely identifies infected individuals in a small number of tests. We discuss the optimal group size and explain why, given the highly infectious nature of the disease, largely parallel searches are preferred. We report proof of concept experiments in which a positive subsample was detected even when diluted a hundred-fold with negative subsamples. Using these methods, the costs of mass testing could be reduced by a large factor. If infected individuals are quickly and effectively quarantined, the prevalence will fall and so will the cost of regular, mass testing. Such a strategy provides a possible pathway to the longterm elimination of SARS-CoV-2. Field trials of our approach are now under way in Rwanda.
\end{abstract}
\maketitle
\end{titlepage}

\section{Introduction}
SARS-CoV-2 represents a major threat to global health. Rapidly identifying and quarantining infected individuals is one of the most important available strategies for containing the virus. However, each diagnostic SARS-CoV-2 test costs 30-50 US dollars~\cite{Medicare2020}. Therefore, testing every individual regularly, as may be required to eliminate the virus, is expensive. The costs are unaffordable for most low-income countries, which have limited available resources for massive SARS-CoV-2 testing. It is therefore important to ask: are there more efficient ways to find infected people? 

The first step in testing, swab collection, is labour intensive but does not require expensive chemicals or equipment. It may therefore be feasible to collect swabs regularly from everyone. The next step involves RT-PCR machines~\cite{Corman2020}. These require expensive chemical reagents, currently in short supply, as well as skilled personnel. Reducing the cost requires that we minimize the total number of tests. Testing rapidly is also vital because SARS-CoV-2 is so infectious. Each RT-PCR test takes several hours in the lab, time during which the virus can spread~\cite{LANL2020}. 

To find infected individuals, the naive approach is to test everyone. This requires one test per person. However, far fewer tests are actually needed, especially at low prevalence where it is much more efficient to pool (or combine) samples and test them together. This idea of group testing dates back to a paper of Dorfman in 1943~\cite{Dorf43} (see \cite{Ellenberg}). Dorfman's algorithm reduces the number of tests per person, required to find all infected individuals, to $\approx 2 \sqrt{p}$ at low prevalence $p$ (see Appendix A). The algorithm we present represents a substantial improvement, requiring only $\approx e p \ln (1/p)$ tests per person at low $p$, where $e=2.718\dots$ is Euler's number.  Measurements in Iceland, early in the pandemic gave $p\approx 0.8\%$~\cite{Iceland}. Recently, in England, following months of lockdown, the prevalence has been estimated (in private-residential households) at a much lower value, $p\sim 0.05-0.12\%$ (95\%CI)~\cite{Pouwels20}. For $p=0.1\%$, Dorfman's strategy yields a cost savings of over 15, while ours achieves over 50. If the prevalence is reduced, these relative cost savings will grow even further. Group testing is therefore rightly attracting renewed interest as countries prepare for widespread SARS-CoV-2 testing in order to contain the outbreak. The Government of Rwanda, in particular, has adopted group testing as a national strategy. All passengers on flights entering or leaving Rwanda must now be tested locally. In order to fly, they must have received a negative test result within the past 72 hours. These tests are being made affordable through group testing. 

Dorfman's algorithm requires two rounds of testing -- the first in which groups are tested, and the second in which every member of every positive group is then tested. Our algorithm involves a similar round of group tests although the optimal group size is larger. The second round consists of ``slicing tests" conducted on the positive groups. Usually, this second round suffices to identify all infected individuals without any individual tests being performed. For less than one positive group in ten, one more round of slicing tests is needed. We shall compare our approach with other approaches in some detail.  There are algorithms which require fewer tests, but which involve many, adaptive rounds of testing. Every extra round consumes significant time, during which the viral prevalence grows. We show that, for this reason, adaptive searches are disfavoured at low prevalence (see Section \ref{sectvi} and Appendix \ref{appendix:a}). There are also non-adaptive algorithms which require, in principle, only one round of testing and hence are faster than ours or Dorfman's. However, they have a significantly higher failure rate for a comparable number of tests (see Appendix~\ref{natm}). 

Group testing is most obviously effective when there are no infected individuals in the group: just one test suffices to show that no-one is infected. Our algorithm takes full advantage of this powerful result. In the first round of tests, subsamples from all group members are pooled and tested together. As we shall show, the optimal group size $N \approx 0.35/p$: the expected number of infected individuals in a group is $0.35$ and a group will test negative over 70\% of the time. Groups that test positive are passed on to the first round of slicing tests, which we now describe. 

\section{When one member of a group is infected } \label{sectii}

Consider the case where only one member of the group is infected. The idea behind our algorithm is geometrical: the group of individuals to be tested is represented by a set of $N$ points on a cubic lattice in $D$ dimensions, organized in the form of a hypercube with $L$ points on a side, so that
\begin{equation}
L^D=N.
\label{e1}
\end{equation}                                                                    
Instead of directly testing the samples taken from every individual, we first divide each of them into $D$ equal subsamples. These $D N$ subsamples are recombined as follows. Slice the hypercube into $L$ planar slices, perpendicular to one of the principal directions on the lattice. Form such a set of slices in each of the $D$ principal directions and pool the $L^{D-1}$ subsamples corresponding to each slice. Altogether, $D L$ slices, each slice combining $N/L =L^{D-1}$ subsamples, are tested in parallel, in a round of slicing tests. If there is one infected individual, then one slice out of the $L$ slices, in each of the $D$ directions, will yield a positive result. That slice indicates the coordinate of the point corresponding to the infected individual, along the associated principal direction. 

\begin{figure}
\centering
\includegraphics[width=0.4\textwidth]{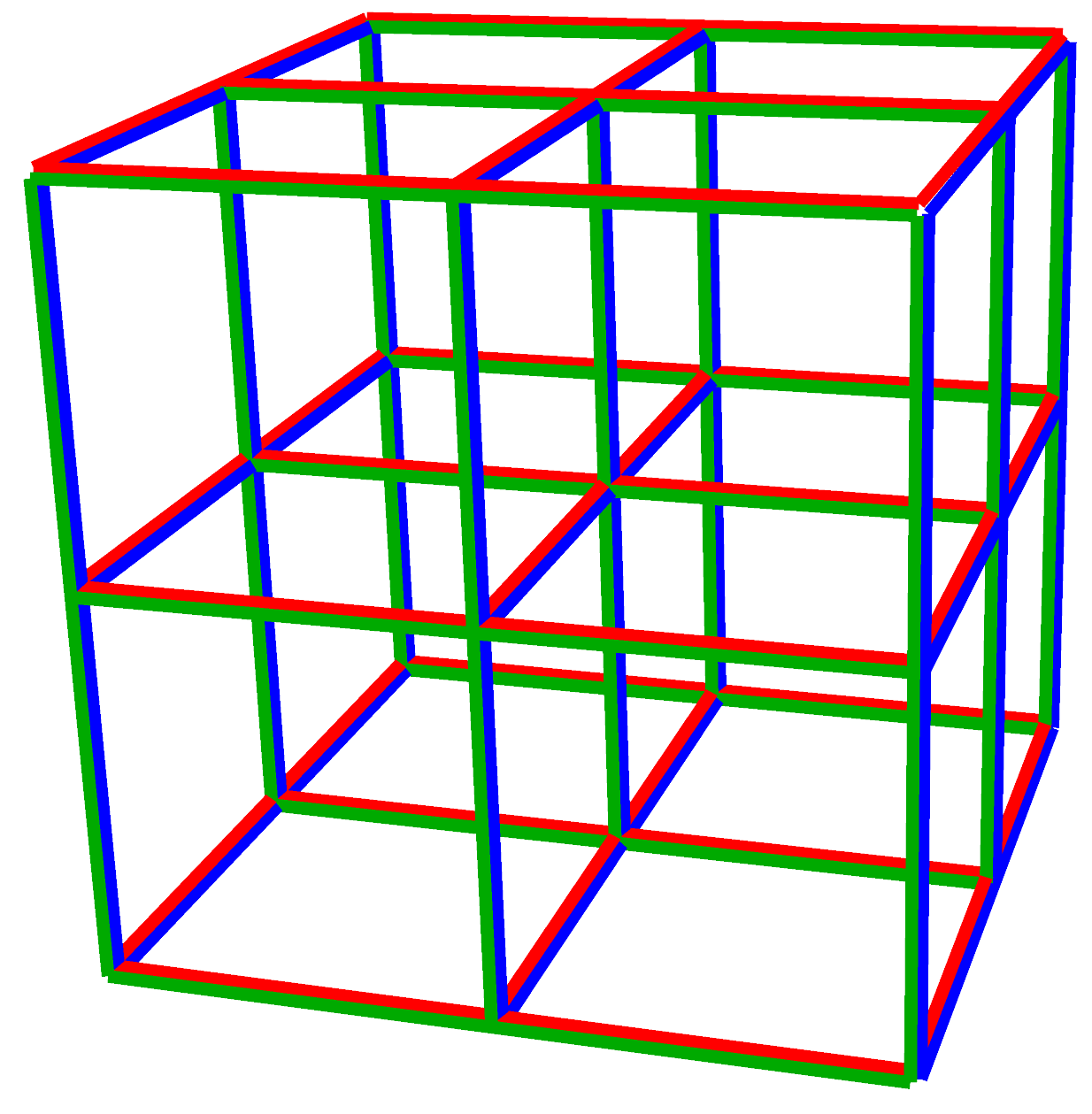}
\caption{Illustration of subsample pooling in the hypercube algorithm, for $D=L=3$ and $N=27=3^3$. Each vertex represents an individual. The hypercube is sliced into $L$ slices, in each of the $D$ principal directions. Samples from $N/L$ individuals are pooled into a sample for each slice. For this example, the 3 sets of slices are shown in blue, red and green. If one infected individual is present, tests on each set of slices identify their coordinate in that direction. Hence only 9 tests uniquely identify them. As the viral prevalence falls, the optimal group size $N$ and dimension $D$ grow, and the efficiency gain rises.}\label{fig:cubes}
\end{figure}

Therefore the number of tests required to uniquely identify the infected individual is
\begin{equation}
D L= D N^{1/D},                             
\label{e2}
\end{equation}                                                                    
where we used (\ref{e1}). Treating $D$ as a continuous variable, the right hand side of (\ref{e2}) diverges at both small and large $D$, possessing a minimum at 
\begin{equation}
D= \ln N ,                             
\label{e3}
\end{equation}                                                                    
corresponding to $L=e$ and a total of $e \ln N$ trials. In reality, $D$ and $L$ must be integers, but using $L=3$ achieves almost the same efficiency (in the total number of trials, $e$ is replaced with $3/\ln3\approx 2.73$, less than half a per cent larger, whereas using $L=2$ or $4$ gives $2/\ln2=4/\ln4\approx 2.89$, more than 5 per cent larger). With no further constraint, finding one infected person in a population of a million, using $L=3$, requires only $39$ tests, performed in one round of testing. (To see this, note that $3^{13} > 10^6$, so a hypercube of side $L=3$ in dimension $D=13$ contains over a million points. A round of slicing tests on this hypercube consists of $D L= 13\times 3 =39$ slice tests.)

\section{Proof of Concept}\label{sectiii}

In practice, we are limited by the capacity of the testing machine. A typical swab yields $10^5$ viral RNA molecules/ml \cite{Wolfel}. For each slice of the hypercube, we combine $N/L$ subsamples of the virus, each of volume $v$. 
If the volume of each combined sample, $V =N v/L$, exceeds the capacity of the PCR we will have to only use a portion of it. We should also keep in mind that at least one viral RNA is needed for an unambiguous result, and we must remain well above this limit. 

Setting $L=3$ and $N=3^D$, we find
\begin{equation}
D = {\ln(V/v)\over \ln 3} + {1}.                               
\label{e4}
\end{equation}      

For example, if $V/v=100$ then (\ref{e4}) yields $D \approx 5$, from which (\ref{e3}) yields $N =243$. If $v$ is a microliter, then $V$ is 100 microliters. In a positive combined sample, there would be $100$ viral RNA molecules. Even if only $10$ microliters are used in the PCR machine, it would contain $10$ viral RNA molecules, sufficient for a positive result. The typical number of tests required to find the infected individual is then only $3 \ln N \approx 17$, a 14-fold improvement in efficiency over naive testing. 

Note that the viral load found in a swab specimen is relatively low if collected during the early stages of viral replication~\cite{Wolfel}. Therefore, swabs taken during this period may contain insufficient virus to yield a positive result. The sensitivity of the test is typically increased by testing specimens collected at sequential time points. Methods like the one we describe here facilitate such sequential testing on a massive scale by drastically reducing the associated costs. In view of the large potential efficiency gains, it is worth exploring whether testing machines could be engineered to accommodate larger test volumes $V$. 

\begin{figure}
\begin{center}
 {\includegraphics[scale=0.6]{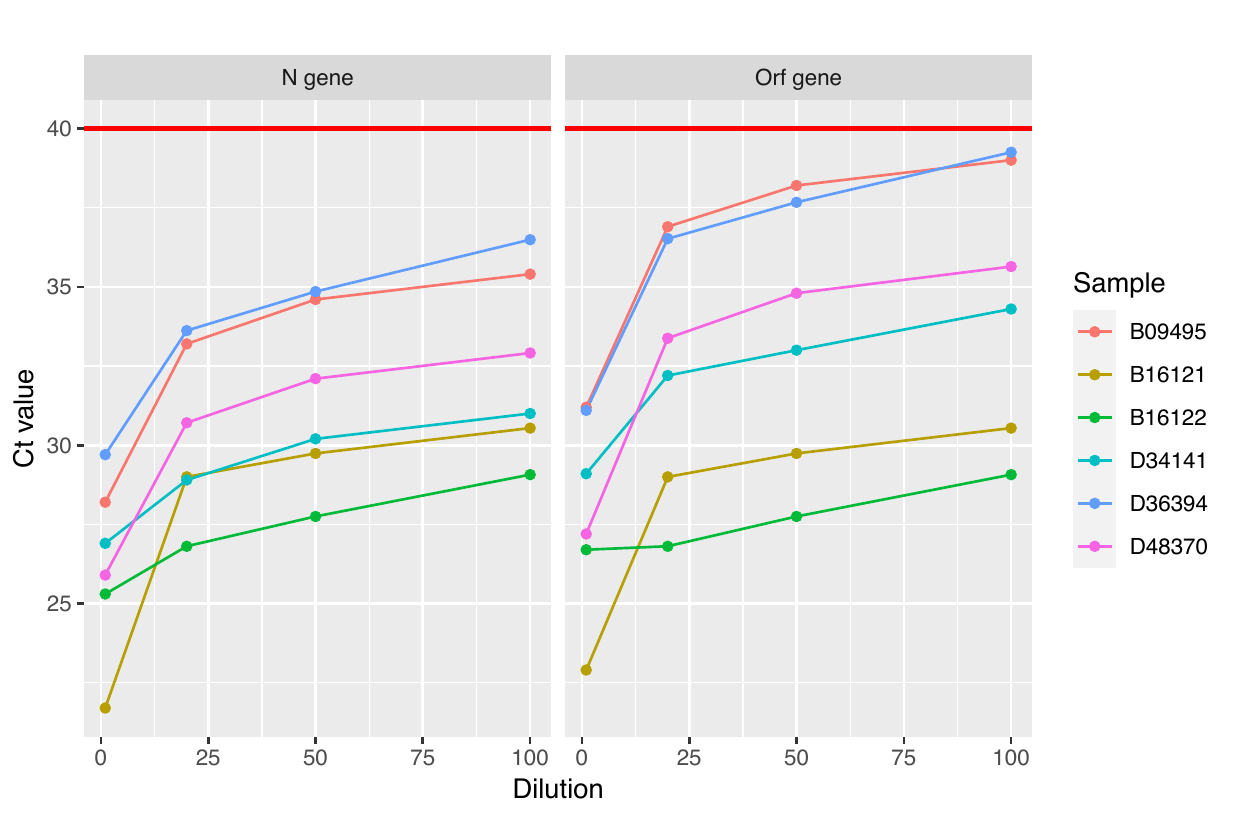}}
\end{center}
\caption{
\textbf{Positive samples are detected after a 100-fold dilution.} \hfill\eject Each of six SARS-CoV-2-positive specimens was diluted through pooling with 19, 49, or 99 negative specimens. A Ct value (i.e. the PCR cycle at which the fluorescence signal generated by a specimen exceeds the baseline signal) was determined for each pool through RT-PCR amplification of the N and Orf genes of SARS-CoV-2. For each gene, the Ct values are plotted (on the y-axis) against the dilution factor (on the x-axis). The red horizontal lines indicate the Ct value (40) at or below which a pool is considered positive. All Ct curves stay below the red lines even as the positive specimens are diluted up to 100-fold. See Appendix C for further experimental details.}
\label{fig2}
\end{figure}

As a proof of concept, using oropharyngeal swab specimens collected during COVID-19 surveillance in Rwanda, we investigated whether known positive specimens still test positive after they are diluted 20-, 50-, or 100-fold ({\it i.e.},  $V/v=20, 50, 100$), through pooling with negative specimens (see Appendix~\ref{methods}). We used a RT-PCR test targetting the N and Orf genes of SARS-CoV-2, which is used routinely for diagnostic screening for COVID-19 in Rwanda. The test is considered to be positive if PCR amplification of both target genes produces an above-background fluorescence signal at a PCR cycle number, {\it i.e.}, a Ct value, that is $\leq 40$. Our key finding  is that positive specimens can still be detected even after they are diluted up to 100-fold (see Fig.~\ref{fig2}). (For recent experiments demonstrating positive sample detection after 30- and 32-fold dilution, see Refs.~\cite{Lohse,Yelin}). As a consistency check, we determined the change in Ct value ($\Delta{Ct}$) that occurs when going from a 50- to a 100-fold dilution. In a perfectly efficient PCR test, the number of target molecules is expected to double after each PCR cycle. Therefore,  the 100-fold diluted positive sample in principle requires one more cycle of PCR amplification than the 50-fold diluted sample, to produce the minimum number of target molecules needed to yield an above-background fluorescence signal. This implies that $\Delta{Ct} \approx 1.0$. Consistent with this expectation, we find that, on average $\Delta{Ct} \approx 1.0$ (s.e. 0.15), for the N gene, and $\Delta{Ct} \approx 1.1$ (s.e. 0.14), for the Orf gene. The changes in Ct values for the other dilutions are consistent with this interpretation. 

Using the measured values of $\Delta{Ct}$ together with the Ct values of a representative sample of positive specimens identified during clinical screening for COVID-19 in Rwanda, we estimated a $95\%$ confidence interval for the Ct values expected after a 20-, 50-, or 100-fold sample dilution (see Appendix~\ref{methods}). Interestingly, for the N gene, we find that a 100-fold dilution produces Ct values with an upper $95\%$ confidence bound of 37.6, below the threshold for positive samples ({\it i.e.} $Ct = 40$). In contrast, for the Orf gene, while the upper $95\%$ confidence bound for Ct values expected after a 50-fold dilution (39.6) is below 40, the corresponding bound for Ct values expected after a 100-fold dilution (40.6) is not. The latter upper confidence bound drops below 40 when the confidence level is reduced to $90\%$. Together, these results indicate that the RT-PCR test used for diagnostic screening for COVID-19 in Rwanda retains a high sensitivity of $\geq 90\%$ if a positive sample is pooled with up to 99 negative samples.

Sample pooling is now used for cost-effective, large scale testing in Rwanda to understand the spatial spread of SARS-Cov-2 nationally, to quickly identify new infections and to enable a rapid response by public health officials. For example, using sample pooling, we recently screened 1,280 individuals using only 64 tests, a 20-fold efficiency gain compared to naive testing (see Figures 5 and 6 in Appendix~\ref{methods}). The cost was thereby reduced to around 3,200 US dollars. Note that, if we had used a single round, non-adaptive testing algorithm (such as the DD algorithm described in Appendix \ref{natm}), assuming a prevalence of 2\%, the total cost (based on the best information rate achieved by the DD algorithm) would have been at least 12,900 US dollars.  By spending ~3 hours checking whether an infected sample was indeed present, more than 9,700 US dollars was saved. Of course, as the number of rounds of testing increases, the cost of delays might eventually offset any accrued savings. The best testing strategy is one that achieves an optimal balance between speed and efficiency.

\section{When more than one member of a group is infected }\label{sectiv}

So far, we have assumed only one member of the group is infected. But what if 2, 3 or more members are infected? In normal circumstances, all we will have is an estimate of the prevalence $p$ of the virus in the population from which the group has been drawn, {\it i.e.}, the probability that a person chosen at random is infected. A feature of group testing is that the first round of group tests, relatively few in number, allows us to conveniently update our knowledge of $p$, even before any infected individuals have been identified (see Appendix \ref{appp}). 

Given $p$, the probability that $k$ members of a group of size $N$ are infected can be described by a Poisson distribution with mean $\lambda = p N$. For $\lambda$ well below unity, the probability falls rapidly with increasing $k$.  As we shall see, this is the regime in which our algorithm operates optimally. At very low prevalence, the optimal $N$ is very large, so $D=\log_L N \gg 1$. The advantages of the hypercube algorithm are particularly obvious in this limit. Therefore, we describe the large $D$ limit first before detailing the algorithm's performance for realistic values of $D$.

The first round of slicing tests, as described in Section \ref{sectii}, yield, for $L=3$, a set of triples of zeros and ones, {\it i.e.}, $\{1,0,0\}$, $\{1,1,0\}$ or $\{1,1,1\}$ and permutations thereof, for every principal direction of the lattice. Let $\sigma$ be the sum of the three values (so $\sigma=1,2$ or $3$) and $d_\sigma$ the number of directions in which the value $\sigma$ occurs, so $d_1+d_2+d_3=D$. For $D\gg1$,  the number of infected individuals $k$ may be accurately inferred from the observed values of $d_\sigma$, even before any infected individuals are identified. Knowing $k$, we then find all infected individuals as follows: 
\begin{enumerate}[label=(\roman*)]
\item If $k=1$, then $d_1=D$. Each positive slice indicates the coordinate of the infected individual in that direction. Hence, the infected individual is identified in one round of slicing tests. 
\item If $k=2$, $d_2>0$ but $d_3=0$. If $d_2=1$, the two infected individuals are immediately identified. If $d_2>1$, we choose one of the directions with $\sigma=2$, and treat the two positive slices as smaller hypercubes, each containing one infected individual. A further round of slicing tests identifies one of them and the other is found by elimination. 
\item  If $k=3$ then, at large $D$, at least one direction has $\sigma=3$. Choose one such direction and treat two of the positive slices as smaller hypercubes, each containing one infected individual. A slicing test on each identifies two infected individuals and the third is found by elimination. 
\item If $k>3$, the number of rounds of slicing tests required to identify all infected individuals is slightly larger than $k$ (see Appendix B).  However, for the optimal value of group size, the probability to have $k>3$ infected members in a group is negligible. 
\end{enumerate}
Hence, in the large $D$ limit, approximately $k$ rounds of slicing tests are required to identify $k$ infected individuals. In Appendix B we compute the expected total number of tests $T$. At low $p$ the number of tests per person is minimised for a group size $N\approx 0.350/p$. At the minimum, $\langle T \rangle/N \approx e p \ln (0.734/p)$ which is plotted as the dashed grey line in Fig.~3. The inverse of this number is the efficiency gain relative to naive testing. 

\begin{figure}
\centering
\includegraphics[width=0.7\textwidth]{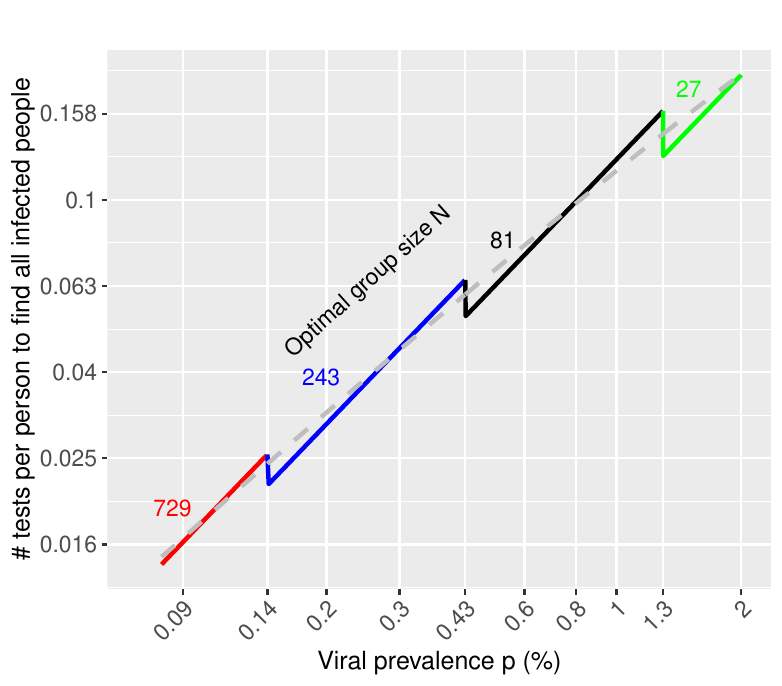}
\caption{The expected number of tests per person to find all infected individuals using the hypercube algorithm (log-log plot). The dashed grey curve shows the result $ e p \ln (0.734/p)$ for the optimal group size $N\approx 0.35/p$, obtained in large $D$ approximation (see Appendix B). The coloured curves show the results obtained from a detailed analysis when the group size $N=3^D$ with $D$ an integer (see Appendices \ref{psok}-\ref{appt} ). Where $0.35/p$ is an exact power of 3, as at the left end of each coloured curve, optimal performance is attained. As $p$ is increased, a growing fraction of sites in the $3^D$ hypercube are left empty, until the next exact power of 3. }
\label{fig:costplot}
\end{figure}

For practical applications, we are interested in the efficiency of the algorithm at modest values of $D$ such as 3, 4 or 5. This requires a more intricate analysis, the technical details of which we relegate to Appendices \ref{psok}-\ref{appt}. However, some simple and general statements may be made. First, when all directions yield $\sigma=1$, only one individual is infected and they are immediately and uniquely identified. This is the most probable outcome of a slicing test. Second, when $\sigma>1$ in only one direction then, if $\sigma=2$, two infected individuals are uniquely identified and for $\sigma=3$ three infected individuals are identified, without the need for any further tests. If $\sigma>1$ in more than one direction, we have to perform a second round of slicing tests. We can reduce the number of required tests by eliminating samples from any slice which tested negative in the first round. In effect, we work with a smaller hypercube, requiring significantly fewer slice tests. We make only one approximation in our analysis, namely we assume the infected samples are rare among the points in the hypercube. It is then an excellent approximation to treat them as independent, randomly chosen points (see Appendix \ref{psok}). Within this approximation, we have computed the probabilities for slice outcomes analytically and followed them through the second round of slice testing.  Strikingly, we find that {\it the hypercube algorithm performs better as $D$ is decreased.} For $\lambda=0.35$ and $D=4$, for example, in $93.3\%$ of cases the slicing tests will end after one round with all infected individuals identified, as explained. For the remaining 6.7\% of cases, one additional round of slicing tests suffices to reduce the failure rate to below $0.01\%$ (see Appendix \ref{appf}). We calculate the expected total number of tests in Appendix \ref{appt}. For $\lambda=0.35$ and $D=4$,  the expected number is 12.3, only marginally higher than $DL$. The expected number of tests per person, for $D=3,4$ and 5, are plotted in  Fig.~3. Where $0.35/p$ is an exact power of 3, as at the left end of each coloured curve, optimal performance is attained. As $p$ is increased, an increasing fraction of sites in the $3^D$ hypercube are left empty, until the next exact power is reached. Nevertheless, pooling still results in a high efficiency gain.\footnote[3]{Further modest efficiency gains could be made by allowing hypercuboids with side lengths of 2 as well as 3.} From Fig.~3, it is evident that the large $D$ approximation provides a surprisingly good (and very convenient) fit to the low $D$ results.

\section{ A worked example}\label{sectv}

It may be helpful to present a worked example, such as might occur in a field trial. Suppose that we want to identify positive samples among 1,200 samples collected from a population in which the prevalence is estimated at $p  = 0.4\%$. We first pool subsamples into 15 groups of 80 members, so the expected number of positive samples in each group, $\lambda=(0.004 \times 1200)⁄15=0.32$, is close to the optimal value of 0.35. We then conduct a single round of 15 tests to identify positive groups. Assuming Poisson statistics, we expect only $27\%$ of the groups to test positive. If we obtain a very different result, we can revise our estimate of $\lambda$ (see Appendix \ref{appp}) and, if necessary, re-group the samples before further testing. Assuming no re-grouping is needed, we place the 80 samples from each of the positive groups in hypercubes with $L=3 $ and $D=4$. In 93\% of cases, one round of slicing tests will suffice to identify all positive samples. In the remaining 7\% of cases, a second round will find the rest, with a failure rate of less than 0.01\%. The expected number of slicing tests per positive group is 12.2.  Hence the total number of tests is $15\times(1+0.27\times12.2)=64.4$, a cost saving, relative to naive testing,  of nearly a factor of 20.   Looking to the future,  if the re-design of RT-PCR machines allows for the pooled testing of larger groups, and if $p$ falls to 0.15\%, the collection of 1,200 samples could be divided into 5 groups of 240 each (corresponding to $\lambda=0.36$). We would then require only $5\times(1+0.30\times15.4)=28.2$ tests to identify the expected 1.8 infected individuals. This represents a cost saving, relative to naive testing, of over 40. 

\section{ Largely Parallel Searches are Preferred}\label{sectvi}

As mentioned in the introduction, there are search methods which require fewer tests. However, they require more rounds of testing. For example, a binary search~\cite{Hwang, Allemann} finds an infected individual in a group of size $N$ in just $\log_2 N$ tests, fewer by a factor of $e \ln 2 \approx 1.88$ than our hypercube algorithm. However, a binary search is adaptive, requiring an iterated sequence of tests.  At low prevalence $p$, the optimal group size $N$ scales as $1/p$ (see Appendix \ref{appendix:b}) so a binary search takes $\sim \log_2 (1/p)$ rounds of testing. For $p=0.4\%$  (or $0.15\%$) a binary search takes 8 (or over 9) rounds of testing whereas a hypercube search takes 2 or at most 3 (see Appendix \ref{appf}). When considering a rapidly spreading infectious disease like COVID-19,  saving time is critically important because infected individuals who are still at large can infect others. A parallel, or largely parallel, strategy such as ours reduces this risk.  Each RT-PCR test takes several hours, to which must be added any time taken for subsample sorting and selection. So, multiple of rounds of testing would consume significant time. During this period, the prevalence $p$ of the virus grows exponentially. The doubling time for the virus is somewhat uncertain but it has been estimated, using data from China~\cite{LANL2020}, to be $\tau_2\approx \,2.4$ days (a recent analysis gives $\tau_2\approx 1.9$ days for New York and $2.9$ days for Los Angeles~\cite{afshordi}). If each RT-PCR test takes $\tau$ days, the viral prevalence grows by $\sim (1/p)^{\tau/ \tau_2}$ during an adaptive search. It follows that, at fixed $\tau/\tau_2$, the adaptive algorithms do worse at low $p$. For example, if we assume that at most 3 rounds of adaptive RT-PCR tests may be performed in a working day, then $\tau=1/3$ day. For prevalences $p$ below $(e \ln 2)^{-\tau_2/\tau} \approx$ 1 per cent, a largely parallel approach like ours is then preferred. Reducing the costs of staff and lab time also favours a largely parallel strategy. Non-adaptive, parallel search algorithms have also been extensively explored in the literature (see Ref.~\cite{Review}). In Appendices   \ref{natm} and \ref{appt}, we show in some detail how the hypercube algorithm outperforms one of the best of these, the DD algorithm with near-constant column weight. Finally, the hypercube algorithm allows for many consistency checks which help to reduce false positives or negatives. For example, if $\sigma=1$ in only one direction, this is very likely to be a false-positive result. In contrast, binary searches rely on repeated testing of the positive sample, making errors harder to identify. 

Implementing the hypercube testing strategy in the lab certainly poses challenges. To mitigate human error, the pooling of samples will need to be automated, specially for the second round of slicing tests where the test must be adapted to a greater extent. Our lab in Rwanda already employs a robot for pooling and we are developing custom software to guide lab procedures. A side benefit will be an ongoing update of the estimated prevalence (see Appendix \ref{appp}), among populations being tested. The incorporation of these mathematical technologies into biological laboratories is a welcome development from which many future benefits may accrue. 

\section{Conclusions} \label{sectvii}

The hypercube algorithm presented here offers, we believe, an attractive compromise between two critically important and competing objectives: minimising the total number of tests to reduce costs and maximising the speed of the testing process to reduce viral spread. We have demonstrated its viability experimentally for group sizes up to 100 and shown that, in principle, one can already achieve cost savings of almost a factor of 20 compared to naive testing (see Fig.~\ref{fig:costplot}), in no more than two rounds of slicing tests, with a failure rate of less than 0.01\%. We pointed out the importance of increasing the sample volume in RT-PCR machines to enable pooled testing of larger group sizes. This could enable even greater cost savings at lower prevalence. In fact, the most striking consequence of our approach is how quickly the cost of testing the whole population falls, advances in RT-PCR machines allowing, with decreasing prevalence. This should incentivise decision-makers to act firmly to test, track and isolate in order to drive the prevalence down. Although mass testing is initially costly, maintaining a low prevalence and, indeed, eliminating COVID-19 will, with the implementation of group testing, become progressively more affordable. 


{\bf Acknowledgments:}
We thank the Rwanda Ministry of Health through RBC for stimulating discussions and correspondence. Kendrick Smith and Corinne Squire provided valuable encouragement and helpful references. Research at AIMS is supported in part by the Carnegie Corporation of New York and by the Government of Canada through the International Development Research Centre and Global Affairs Canada. Research at Perimeter Institute is supported in part by the Government of Canada through the Department of Innovation, Science and Economic Development Canada and by the Province of Ontario through the Ministry of Colleges and Universities.

\noindent {\bf Contributions:} 
 L.M., S.N. coordinated the experiments; P.N., T.N., E.N., M.S., C.M., D.N., M.N. contributed to patient recruitment and data collection from the community; P.N., Y.B., J.S., A.U., R.R., E.L.N., E.M., E.M., N.R., I.E.M., J.B.M., C.M.M. and E.U. contributed to laboratory RT-PCR test validation, data analysis and interpretation. W.N. and N.T. contributed to the theory.

\bibliographystyle{utphys}
\bibliography{library}

\providecommand{\href}[2]{#2}\begingroup\raggedright\begin{thebibliography}{10}

\bibitem{Medicare2020}
{Medicare}, ``{Medicare administrative contractor (MAC) COVID-19 test
  pricing},''.
  \url{https://www.cms.gov/files/document/mac-covid-19-test-pricing.pdf}.

\bibitem{Corman2020}
{V.M.~Corman, O.~Landt, M.~Kaiser {\it et al.} }, ``{Detection of 2019 novel
  coronavirus (2019-nCoV) by real-time RT-PCR},'' {\em Euro. Surveill.} {\bf
  25} (2020)  2000045.

\bibitem{LANL2020}
{S.~Sanche, Y.T.~Lin, C.~Xu, E.~Romero-Severson, M.~Hengartner, R.~Ke}, ``{High
  contagiousness and rapid spread of severe acute respiratory syndrome
  coronavirus},'' {\em {Emerg. Infect. Dis.}} {\bf 26} ({July, 2020})  .
  \url{https://doi.org/10.3201/eid2607.200282}.

\bibitem{Dorf43}
{R.~Dorfman}, ``{The Detection of Defective Members of Large Populations},''
  {\em Ann. Math. Stat.} {\bf 14} (1943)  436.

\bibitem{Ellenberg}
J.~Ellenberg, ``{Five people. One test. This is how you get there.},'' {\em
  {New York Times}} (May 7, 2020)  .
  \url{https://www.nytimes.com/2020/05/07/opinion/coronavirus-group-testing.html}.

\bibitem{Iceland}
{D.F.~Gudbjartsson {\it et. al.}}, ``{Spread of SARS-CoV-2 in the Icelandic
  Population},'' {\em New Eng. J. Med.} {\bf {DOI: 10.1056/NEJMoa2006100}}
  (April 14, 2020)  .
  \url{https://www.nejm.org/doi/full/10.1056/NEJMoa2006100}.

\bibitem{Pouwels20}
K.~B. Pouwels, T.~House, {\em et al.}, ``Community prevalence of sars-cov-2 in
  england: Results from the ons coronavirus infection survey pilot,''
  \href{http://dx.doi.org/10.1101/2020.07.06.20147348}{{\em medRxiv} (2020)  },
  \href{http://arxiv.org/abs/https://www.medrxiv.org/content/early/2020/07/07/2020.07.06.20147348.full.pdf}{{\tt
  https://www.medrxiv.org/content/early/2020/07/07/2020.07.06.20147348.full.pdf}}.

\bibitem{Wolfel}
{R.~W\"olfel, V.M.~Corman, W.~Guggemos {\it et al.} }, ``{Virological
  assessment of hospitalized patients with COVID-2019},'' {\em {Nature}}  .
  \url{https://doi.org/10.1038/s41586-020-2196-x}.

\bibitem{Lohse}
{S.~Lohse {\it et. al.}}, ``{Pooling of Samples for testing for SARS-CoV-2 in
  asymptomatic people},'' {\em Lancet Infectious Diseases} (April 28, 2020)  .
  \url{https://doi.org/10.1016/ S1473-3099(20)30362-5}.

\bibitem{Yelin}
{I.~Yelin {\it et. al.}}, ``{Evaluation of COVID-19 RT-qPCR test in
  multi-sample pools},'' {\em Clinical Infectious Diseases} (May 02, 2020)  .
  \url{https://doi.org/10.1093/cid/ciaa531}.

\bibitem{Hwang}
{F.K.~Hwang}, ``{A method for detecting all defective members in a population
  by group testing},'' {\em J. Amer. Stat. Assoc.} {\bf 67} (1972)  605.

\bibitem{Allemann}
{A.~Allemann}, ``{An efficient algorithm for combinatorial group testing},''
  {\em {{\rm in Aydinian H., Cicalese F., Deppe C. (eds)}, Information Theory,
  Combinatorics, and Search Theory, {\rm Lecture Notes in Computer Science,
  {\bf 7777}, Springer, 2013}}}  .

\bibitem{afshordi}
N.~Afshordi, B.~Holder, M.~Bahrami, and D.~Lichtblau, ``Diverse local epidemics
  reveal the distinct effects of population density, demographics, climate,
  depletion of susceptibles, and intervention in the first wave of covid-19 in
  the united states,'' \href{http://arxiv.org/abs/2007.00159}{{\tt
  arXiv:2007.00159 [q-bio.PE]}}.

\bibitem{Review}
O.~J. M.~Aldridge and J.~Scarlett, ``Group testing: an information theory
  perspective,'' {\em Foundations and Trends in Communications and Information
  Theory, to appear}  , \href{http://arxiv.org/abs/1902.06002}{{\tt
  arXiv:1902.06002}}. \url{http://arxiv.org/abs/1902.06002}.

\bibitem{Williams}
B.~G. Williams, ``{Optimal pooling strategies for laboratory testing},''
  \href{http://arxiv.org/abs/1007.4903}{{\tt arXiv:1007.4903 [q-bio.QM]}}.

\bibitem{Eliaz}
M.~D. Y.~Eliaz and G.~Gasic, ``{Poolkeh Finds the Optimal Pooling Strategy for
  a Population-wide COVID-19 Testing (Israel, UK, and US as Test Cases)},''
  \href{http://dx.doi.org/10.1101/2020.04.25.20079343}{{\em medRxiv} (2020)  }.
  \url{https://www.medrxiv.org/content/early/2020/05/05/2020.04.25.20079343}.

\bibitem{Shani-Narkiss}
N.~Y. H.~Shani-Narkiss, O.D.~Gilday and I.~Landau, ``{Efficient and Practical
  Sample Pooling for High-Throughput PCR Diagnosis of COVID-19},''
  \href{http://dx.doi.org/10.1101/2020.04.06.20052159}{{\em medRxiv} (2020)  }.
  \url{https://www.medrxiv.org/content/early/2020/04/14/2020.04.06.20052159}.

\bibitem{Lakdawalla}
D.~G. D.~Lakdawalla, E.~Keeler and E.~Trish, ``{Getting Americans back to work
  (and school) with pooled testing},'' {\em {USC Schaeffer Center White Paper}}
  (2020)  .
  \url{https://healthpolicy.usc.edu/wp-content/uploads/2020/05/USC_Schaeffer_PooledTesting_WhitePaper_FINAL-1.pdf}.

\bibitem{ABJ}
{M.~Aldridge, L.~Baldassini, and O.T.~Johnson}, ``{Group testing algorithms:
  Bounds and simulations},'' {\em {IEEE Transactions on Information Theory}}
  {\bf {60(6)}} (2014)  , \href{http://arxiv.org/abs/{1306.6438}}{{\tt
  arXiv:{1306.6438}}}. \url{http://arxiv.org/abs/1306.6438}.

\bibitem{ONSmethods}
{Office of National Statistics, UK}, ``Covid-19 infection survey (pilot):
  methods and further information.''
  \url{https://www.ons.gov.uk/peoplepopulationandcommunity/healthandsocialcare/conditionsanddiseases/methodologies/covid19infectionsurveypilotmethodsandfurtherinformation#test-sensitivity-and-specificity},
  23, July, 2020.

\bibitem{ghosh2020compressed}
S.~Ghosh, R.~Agarwal, M.~A. Rehan, S.~Pathak, P.~Agrawal, Y.~Gupta, S.~Consul,
  N.~Gupta, R.~Goyal, A.~Rajwade, and M.~Gopalkrishnan, ``A compressed sensing
  approach to group-testing for covid-19 detection,''
  \href{http://arxiv.org/abs/2005.07895}{{\tt arXiv:2005.07895 [q-bio.QM]}}.

\bibitem{Daan}
{DAAN Gene Company Ltd.}, ``Daan - rt - pcr reagent set for covid-19.''
  \url{https://prolabcorp.com/daan-rt-pcr-reagent-set-for-covid-19-real-time-detection-for-48-samples-research-use-only/},
  May, 2020.

\bibitem{WHO}
{World Health Organisation}, ``Sars‐cov‐2 nucleic acid tests.''
  \url{https://www.who.int/diagnostics_laboratory/200414_eul_covid19_ivd_update.pdf?ua=1},
  May, 2020.

\end{thebibliography}\endgroup
\appendix

\section{Comparison with Dorfman}
\label{appendix:b}

In a landmark paper in 1943, R. Dorfman considered the problem of searching for infected individuals by grouping (or pooling) samples. His approach remains influential (see, {\it e.g.}, Refs.~\cite{Williams,Eliaz,Shani-Narkiss,Lakdawalla}). Consider a population of $n$ individuals, broken up into groups of $N$ members each. If the probability that any individual is infected (the prevalence) is $p$, the probability that a group is free of infection is $(1-p)^{N}$. Conversely, the probability that at least one member is infected is $p'=1-(1-p)^{N}$. Dorfman's strategy was to test all groups, and then to test every member of every infected group. The expected number of tests is then
\begin{equation}
     \langle T \rangle   = {n\over N}+p' n,                
\label{a1}
\end{equation}      
and the number of tests required per person is
\begin{equation}
     {\langle T \rangle\over n}   = {1\over N} +p' =   {1\over N}+ 1-(1-p)^{N}   \approx {1\over N}                +p {N}.
\label{a2}
\end{equation}      
In the last step, we assumed that $p\ll 1$. The number of tests per person is minimised when the group size $N=1/\sqrt{p}$. The expected number of tests per person, at the optimal group size, is approximately $2\sqrt{p}$. 

Let us compare these results with those obtained using our hypercube algorithm at large $D$. Assuming Poisson statistics for the number of infected individuals $k$ in a group of size $N$, with $\lambda=p N$, the expected number of tests per person is given by
\begin{eqnarray}
     {\langle T \rangle\over N}   &=& {e^{-\lambda} \over N}\left(1 +e \ln N \sum_{k=1}^\infty {\lambda^k \over k!} r_k \right)
 \approx {e^{-p N} \over N} + p e \ln N 
     \label{a3}
\end{eqnarray}     
where $r_k$ denotes the number of rounds of slicing tests, and $e\ln N$ is the number of tests in each round as explained in Section \ref{sectii}. In the second,  approximate equation, we have set $r_k\approx k$. For $k=1-3$ this is exact, as explained in Section \ref{sectiv}. For $k>3$ more than $k$ rounds may be needed. Let $r_k=k+c_k$ where $c_k$ is the average excess number of rounds of slicing tests (and is nonzero only for $k>3$). These numbers are tedious but straightforward to compute. The first few are: $c_4={1\over 3}$, $c_5={16\over 33}$, $c_6={37\over 54}$. In any case, they turn out to be unimportant because the minimum of (\ref{a3}) is dominated by contributions from lower $k$. Note that, in the large $D$ approximation, we ignore any changes in the argument of the logarithm due to fewer than $N$ tests sometimes being needed. These changes are unimportant at large $D$ but become important at lower $D$ and are carefully taken into account in Appendices \ref{psok}-\ref{appt}. 

The last expression in (\ref{a3}) represents the approximation that the $c_k$ corrections are neglected. The first term diverges at small $N$, and the second diverges at large $N$. Thus, a minimum exists. It is located at $\lambda =p N\approx 0.350$, independent of $p$. For such a small value of $\lambda$, contributions from  $k>3$ are strongly suppressed. One subtlety is that, at very low $p$, since $N=\lambda/p$, for the relevant values of $\lambda$ the argument of the logarithm becomes very large, so its derivative with respect to $\lambda$ becomes suppressed with respect to its value. This effect means that, for extremely small $p$, the derivative of the corrections can compete with the derivative of the logarithm. We have checked that only for $p<10^{-10}$ does this alter the minimum value of $\lambda$ by more than $0.01$. Hence, for all practical purposes, we can safely ignore the $c_k$ corrections. 

The optimal group size and the corresponding minimal number of tests per person are thus given, to an excellent approximation, by
\begin{equation}
  N\approx {0.350\over  p}; \qquad \quad   {\langle T \rangle\over N}   \approx e p \ln \left({0.734\over p}\right) .
\label{a4}
\end{equation}   
For a prevalence $p$ of 1 per cent, Dorfman's approach yields a group size of approximately 10 and approximately 0.2 tests per person, whereas ours yields a group size of 35 and 0.12 tests per person. For a prevalence of 0.1 per cent, Dorfman's optimal group size is 32 whereas ours is 350. His approach requires 0.06 tests per person, whereas ours needs only 0.018. For all lower, but still reasonable, values of $p$ our approach prevails by an increasing margin. 

Further geometrical insight into our approach and its relation to Dorfman's may be gained by considering its generalization to a hypercuboid. A set of $n$ people to be tested may be represented as a rectangular volume in a $D$-dimensional cubic lattice with $L_1, L_2, \dots, L_D$ points on a side, constrained to obey $L_1 L_2\cdots L_D=n$. For simplicity, consider the case when there is exactly one infected individual to be found. Our approach is to group the points in slices taken along each of the principal directions of the lattice, and to test each slice. The required number of tests is $L_1+L_2+\dots +L_D$. This is minimized, for fixed $n$, when the $L_i$ are all equal, {\it i.e.}, for a hypercube. One can think of Dorfman's approach to the same problem as a reduced $D=2$ version, with $L_1$ the number of groups and $L_2$ the group size so $L_1 L_2=n$.  Dorfman's approach is first to test the $L_1$ groups, and then the $L_2$ members of the positive group (this is the sense in which it is reduced: in the second step, he tests individuals not groups). The number of tests required is $L_1+L_2$ which is minimized, at fixed $n=L_1 L_2$, by taking $L_1=L_2=\sqrt{n}$ so this is the optimal group size. Setting the prevalence $p=1/n$, we recover the results noted for Dorfman's algorithm above. The advantage of our approach over Dorfman's is that of going to higher dimensions.  Similarly, in other testing algorithms, test matrices are designed using arguments which are essentially two dimensional~\cite{Review} where in our approach, the higher dimensional viewpoint is key.

\section{Information theory bounds}
\label{appendix:a}
Information theory sets a lower bound on the number of tests required to uniquely identify all infected individuals. The uncertainty in who is infected is associated with an entropy, 
\begin{equation}
    S=-\sum_i p_i \ln p_i,      
\label{a11}
\end{equation}   
where the sum is over all possible states and the $p_i$ are the corresponding probabilities. If a test outputs a zero or a one then for $t$ tests the number of possible test outputs is $2^t$ and the corresponding information gained is at most $t \ln 2$. In order to learn everything about the system, one requires an information gain of at least $S$, hence  
\begin{equation}
  t> {S\over \ln 2}.
\label{a12}
\end{equation}   
Consider a sample of size $n$, with $k$ infected individuals chosen at random. The number of such states is ${n \choose k}$. Therefore, from (\ref{a12}), the minimum number of tests required is $\log_2{n \choose k}\sim k \log_2(n/k)$ for $k\ll n$.  Assuming a binomial distribution with prevalence $p$, and replacing $k$ with its expectation $p\, n$, we find the expected number of tests per person is $\sim p \log_2(1/p)$.  Binary searches can approach this limit by performing an adaptive series of tests (see, {\it e.g}, Refs.~\cite{Hwang, Allemann}). However, binary searches are adaptive, requiring an iterated series of tests. The number of rounds of testing  $\sim \log_2 n \sim \log_2 (1/p)$ for the optimal value of $n$, at low $p$. Parallel searches, called ``noiseless, nonadaptive'' tests in Ref.~\cite{Review}, are faster but require more tests as we discuss in Appendix \ref{natm}. 

\section{Comparison with non-adaptive test matrix searches}\label{natm}

Proposals for group testing have been extensively explored in the computer science literature - see Ref.~\cite{Review} for an excellent recent review. The ``noiseless, non-adaptive" tests described in Chapter 2 there are most relevant. Our hypercube algorithm should be viewed as semi-adaptive because it sometimes takes two rounds of slicing tests to achieve a very low failure rate. As we shall explain, the best non-adaptive tests discussed in Ref.~\cite{Review} are not directly applicable to the problem considered here. These methods rely on building a $T$ by $N$ ``test matrix" of zeros and ones, where $T$ is the number of tests and $N$ is the number of samples to be tested. When applied to a column vector of candidate samples, each row of the matrix selects those which are to be pooled in the corresponding test. The zeros and ones of the test matrix are chosen at random according to some probability distribution. In the simplest case this is a Bernoulli distribution, with $p$ being the probability to be one and $1-p$ the probability to be zero. An important improvement is known as the ``near-constant column weight" design. The idea is to ensure that every sample participates in a similar number of tests, by choosing  a fixed number of entries (called the weight) of each column uniformly at random with replacement and setting them to one, while setting all other elements zero. The weight, which in practice provides an upper bound to the number of nonzero column entries (because of the choice with replacement - hence the term `near-constant') is adjusted to optimise the search. 

With the test matrix fixed, the next step is to interpret the test outcomes. One of the best-performing algorithms is the ``definitely defective" or DD algorithm due to Aldridge, Baldassini and Johnson~\cite{ABJ}. The terminology arises from Dorfman's original paper where infected samples were termed ``defective." The DD algorithm proceeds in two steps: a) eliminate all samples which were included in negative tests (because they are definitely non-defective) and label the remainder as ``possibly defective" and b) identify all samples which are the {\it only} ``possibly defective" samples included in a positive test, as ``definitely defective."  The DD algorithm outputs a set of definitely defective samples which may or may not be the complete set of defectives. If some defectives have been missed, the algorithm has failed. The DD algorithm with near-constant column weight design performs particularly well in simulations when the weight is suitably optimised. The algorithm is also tractable analytically in the limit where the number of defectives $k \sim \Theta(n^\alpha)$ and $n\rightarrow \infty$ with $\alpha$ a constant. For small $\alpha$, the DD algorithm with near-constant column weight design has a rate of at least $\ln 2\approx 0.693$ bits (see Theorem 2.8 and Eq.~(2.18) in Ref.~\cite{Review}), where the ``rate" of an algorithm is defined to be the information gained per test, in bits. While these results are instructive they are, unfortunately, not directly relevant. As the authors of \cite{Review} explain, the optimal weight for the DD algorithm is the closest integer to $(T/k)\ln 2$ where $T$ is the number of tests and $k$ is the number of defectives. {\it So $k$ must be known in advance}, which is an unrealistic requirement in our context. Second, we are interested in optimal performance at fixed prevalence $p$, not fixed $\alpha$. With these caveats, it is interesting nevertheless that the information rate achieved by the DD algorithm with near-constant column weight design and in the analytical limit they discuss, is larger by a factor of $e (\ln 2)^2\approx 1.3$ than that achieved by our method at large $D$. Equally interesting, we have found the performance of the hypercube algorithm to improve as $D$ decreases. In Appendix \ref{appt}, we show that the``rate" (in the sense of \cite{Review}) achieved by the hypercube algorithm {\it with no prior knowledge of $k$} other than the prevalence $p$ which is used to set the group size, increases with decreasing $D$. For $D=3$ it actually exceeds the above mentioned ``rate" of $ 0.693$ bits.  

Another key performance indicator is a test's failure rate. To compare the hypercube algorithm with the DD algorithm with near-constant column weight design, we have implemented a numerical code for the latter. Our code accurately reproduces the performance as plotted in Figure 2.3 of Ref.~\cite{Review}, on a test problem with $k=10$ and $n=500$. We then used the same code to study the success rate for the following test problem, chosen to correspond as closely as possible to the problem of interest here but still be addressable using their algorithm. First, we consider groups of size $N$, with a single infected member. (This is unrealistic, of course, but allows us to select the optimal weight for the DD algorithm, and ensures the hypercube algorithm always succeeds). For $N=81$, the hypercube algorithm requires only 12 tests to identify the infected sample. In contrast, we find the DD algorithm on average requires more than 18 tests to achieve a failure rate below 0.1\%\footnote[3]{It would be unreasonable, in our view, to use a search algorithm which introduced theoretical errors comparable in magnitude to the smallest experimental error because analysis is cheap compared to experiment. In the case of RT-PCR tests, the smallest error is the probability for false positives, which is known to be  $<0.1\%$\cite{ONSmethods}. We insist that search method errors should be smaller.} Likewise, for $N=243$, our algorithm requires 15 tests whereas the DD algorithm requires more than 21 tests to achieve a failure rate below 0.1\%. For this problem, both algorithms take only one round of testing, so they are equally fast.  Next, we examine the more realistic situation, where only the prevalence is known. Setting $p=0.35/N$, we generate a set of randomly chosen groups of size $N$, having a Poisson distribution for the number $k$ of infected members. In Appendix \ref{appt},  we show that the hypercube algorithm requires at most two rounds of slicing tests to achieve a failure rate below $0.01\%$. When it fails, it informs us that it has failed so that, if necessary, we can run further tests. However, the DD algorithm can fail to identify a positive sample (because it is not yet ``definitely defective"), without providing any indication of this failure. For the DD algorithm applied to a Poisson distribution of infected groups, if we choose the column weight to be that for the mean value of $k$, we find a large deterioration in its performance, with a much higher failure rate. This is not surprising since, when $k$ deviates from the mean, as it usually does, the column weight is sub-optimal. Finally, let us also mention the recent work of Ref.~\cite{ghosh2020compressed}. From the data they present, their algorithm produces false positives at a rate which exceeds the experimental error.$^\ddagger$ There are other significant differences with our approach - for example, they use continuous data from the RT-PCR tests.   It will be interesting to compare the efficiency and failure rates of their algorithm with those of ours on test problems and we intend to do so in the near future. 

For the problem at hand, we conclude that the semi-adaptive hypercube search algorithm has multiple advantages over the DD algorithm with near-constant column weight design. These arise from its formulation in higher dimensions. 

\section{Probabilities of slice outcomes given $k$}\label{psok}

As discussed in Section \ref{sectiv}, the output of the first round of slicing tests on positive groups is, for $L=3$, a set of triples with $\sigma=1,2$ or 3. Given a population of interest (more precisely, a sample of individuals drawn from the population) and an assumed prevalence $p$, we form groups of size $\approx 0.35/p$ for hypercube screening. It is convenient to compute the probabilities for obtaining each value of $\sigma$ in any given hypercube direction, in a ``dilute gas" approximation. Namely, we assume that infected samples are each placed at a random point in the hypercube corresponding to the group. This approximation ignores the constraint that no two infected samples may occupy the same point. But since this circumstance is very rare at low viral prevalence, at even modest values of $D$, the approximation is excellent. 
In this approximation, if there are $k$ infected individuals in the population, the probability that an infected sample is placed in a given slice is just $1/3$. Let $P_L(\sigma |k)$ be the probability of obtaining $\sigma$ given $k$ (at fixed $L$). Then
\begin{equation}
P_3(1|k)= \binom{3}{1} {1\over 3^k} \sum_{\substack{l,m=0 \\ l+m+n=k}} \binom{k}{l,m,n} = 3 {1\over 3 ^k}.
\label{pso1}
\end{equation}   
Similarly, 
\begin{equation}
P_3(2|k)= \binom{3}{2} {1\over 3^k} \sum_{\substack{l,m>0 \\ l+m=k}} \binom{k}{l} = 3{1\over 3^k} \left[\sum_{l=0}^k \binom{k}{l} -2\right] =6(2^{k-1} -1) {1\over 3^k},
\label{pso2}
\end{equation}   
and finally,  
\begin{eqnarray}
P_3(3|k)&=& 1- P_3(1|k)-P_3(2|k) =3(3^{k-1}-2^k +1) {1\over 3^k}
\label{pso3}
\end{eqnarray}   

In subsequent Appendices, we shall need these formulae both for estimating the viral prevalence or for identifying infected group members. In particular, we shall need conditional probabilities where we make use of the results of previous tests. In this case, we need to normalize the probabilities for the allowed values of $\sigma$ to ensure they sum to unity. So, for example, if we know that $\sigma\neq 3$ in any direction, then we obtain;
\begin{eqnarray}
P_3(1|k,d_3=0)&=& {1\over 2^k-1}; \quad P_3(2|k,d_3=0)= 1-{1\over 2^k-1}.
\label{pso4}
\end{eqnarray}    
Also, in subsequent rounds of slicing tests, we need the probabilities for smaller hypercubes with $L=2$,
\begin{equation}
P_2(1|k)= {1\over 2^{k-1}}; \qquad P_2(2|k)= 1-{1\over 2^{k-1}}.
\label{pso5}
\end{equation}

\section{Probabilities of slice outcomes given $\lambda$}\label{psol}

Typically, when a positive group is detected, we will not know $k$, the number of infected samples it contains. We will, however, have available the updated estimate of the number of infected individuals expected in the hypercube, $\lambda^{(1)}$ (see the first paragraph in Appendix \ref{appp}). This allows us to compute probabilities for the results of the first round of slicing tests. For the second round of slicing, we will subdivide the original hypercube into a number of smaller ones. We shall re-estimate the number of infected individuals in these smaller hypercubes, $\lambda^{(2)}$, in the light of the first round of slicing test results, and so on. 

Let  $P_L(\sigma|\lambda^{(1)})$ be the probability of obtaining $\sigma$ in any given hypercube direction, in the first round of slicing tests, in any group. Since only positive groups are tested, for any given group we have
\begin{equation}
P_3(1|\lambda^{(1)})= \sum_{k=1}^\infty P_3(1|k)P(k|\lambda^{(1)})=\sum_{k=1}^\infty {1\over 3^{k-1}}{e^{-\lambda^{(1)}} \over 1-e^{-\lambda^{(1)}} }{(\lambda^{(1)})^k\over k!} ={3 \over 1+e^{\lambda^{(1)}/3}+e^{2\lambda^{(1)}/3}},
\label{psol1}
\end{equation}   
where, because of the restriction $k>0$,  $P(k|\lambda^{(1)})$ is a truncated Poisson distribution hence the additional factor of $1-e^{-\lambda^{(1)}}$ in the denominator. 
Similarly, 
\begin{equation}
P_3(2|\lambda^{(1)})={3 (e^{\lambda^{(1)}/3}-1)\over 1+e^{\lambda^{(1)}/3}+e^{2\lambda^{(1)}/3}},
\label{psol2}
\end{equation}   
and 
\begin{equation}
P_3(3|\lambda^{(1)})={(e^{\lambda^{(1)}/3}-1)^2\over 1+e^{\lambda^{(1)}/3}+e^{2\lambda^{(1)}/3}},
\label{psol3}
\end{equation}   
satisfying $\sum_{\sigma=1}^3P_3(\sigma|\lambda^{(1)})=1$. It will be convenient in what follows to also have the probabilities for $\sigma=1$ and $\sigma=2$, conditioned on $d_3=0$ or $d_3=1$. For example, making use of (\ref{pso4}) we have
\begin{equation}
P_3(1|\lambda^{(1)},d_3=0)={e^{-\lambda^{(1)}} \over 1-e^{-\lambda^{(1)}} }\sum_{k=1}^\infty {1\over 2^{k}-1}{(\lambda^{(1)})^k\over k!}; \, P_3(2|\lambda^{(1)},d_3=0)=1-P_3(1|\lambda^{(1)},d_3=0).
\label{psol4}
\end{equation}   

For understanding the failure rate of our algorithm in subsequent rounds of slicing tests, $j>1$, we will need the probabilities of obtaining $\sigma=1$ or $2$ in a smaller hypercube of length $L=2$. 
Using (\ref{pso5}), these are given by
\begin{equation}
P_2(1|\lambda^{(j)})=\sum_{k=1}^\infty {1\over 2^{k-1}}{ e^{-\lambda^{(j)} } \over 1-e^{-\lambda^{(j)} } }{(\lambda^{(j)})^k\over k!}={2\over e^{\lambda^{(j)}/2 }+1}; \quad P_2(2|\lambda^{(j)})=1-P_2(1|\lambda^{(j)}),
\label{psol6}
\end{equation}   
where we again renormalised the Poisson distribution because $k>0$ in each of the smaller hypercubes. We compute the relevant values of $\lambda^{(j)}$ in Appendix \ref{appf}.  From the dependence on $\lambda^{(j)}$ it should always be clear to which round of testing the probabilities apply. 

\section{Expected Failure Rate }
 \label{appf}
 
If, after the first round of slicing tests, $\sigma>1$ in more than one hypercube direction, we will be unable to uniquely identify the coordinates of all infected samples. The failure rate for our algorithm in this first round is the corresponding probability, which may be expressed as one minus the probability that $\sigma=1$ in all but one direction, namely 
 \begin{equation}
F^{(1)}=1-\sum_{d=D-1}^D {\rm Pr}(d,D,P(1|\lambda^{(1)})),
\label{efr1}
\end{equation}   
where ${\rm Pr}(d,D,p)\equiv p^d (1-p)^{D-d} D!/(d!(D-d)!) $ is the binomial probability distribution and $P(1|\lambda^{(1)})$ is given in (\ref{psol1}). 

To compute the failure rate in subsequent rounds of slicing tests, it is helpful to separate the possible outcomes of the first test into three classes: $d_3=0$, $d_3=1$ and $d_3>1$. For realistic values of $D$, and $\lambda\approx 0.35$,  the outcomes with $d_3=0$ strongly predominate, followed by those with $d_3=1$ which in turn predominate over higher values of $d_3$. More precisely, 
 \begin{equation}
 {\rm Pr}(0,D,P_3(3|\lambda^{(1)}))\gg {\rm Pr}(1,D,P_3(3|\lambda^{(1)}))\gg \sum_{d=2}^D {\rm Pr}(d,D,P_3(3|\lambda^{(1)})),
\label{efr2}
\end{equation}   
for all practical values of $D$ and $\lambda^{(1)}$. For example, for $\lambda^{(1)}=0.35$ and $D=4$, of the first round of slicing tests,  $98.2\%$ give $d_3=0$, $1.78\%$ give $d_3=1$ and only $0.012\%$  give $d_3>1$. (Even for $D=10$, the probability that $d_3>1$ is less than $0.1\%$.) The strong dominance of $d_3=0$ and $d_3=1$ over other values allows us to focus our effort on these cases.  

If $d_3=0$, all of the positive sample coordinates reside in a smaller hypercube with volume $2^{d_2}$. We divide it in two along any one of its principal directions, forming two hypercubes of dimension $d_2-1$. Should we require further rounds of testing, this subdivision will be iterated so that the $j$'th round of testing will involve $2^{j-1}$ hypercubes of length $L=2$. Because the positive samples are distributed randomly among the hypercubes, in the $j$'th round the number of 
positive samples expected in each smaller hypercube is 
\begin{equation}
 \lambda^{(j)}={ \lambda^{(1)}\over 2^{j-1}}
\label{efr3}
\end{equation}   

\begin{figure}
\centering
\includegraphics[width=1\textwidth]{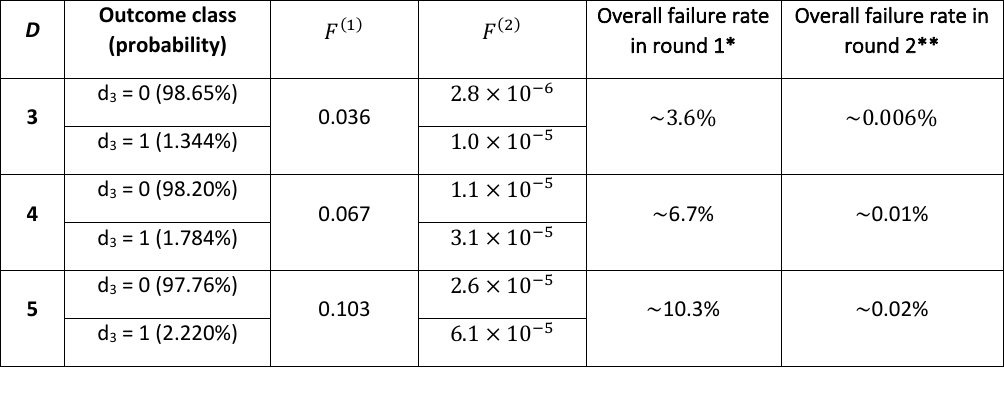}
\caption{ {\bf Failure rates} for $D=3,4,5.$ When $\lambda^{(1)}=0.35$, one round of slicing tests suffices to find all infected samples in $\approx 96.4\%$, $93.3\%$ and $89.7\%$ of cases, for $D=3, 4$ and $5$ respectively. Hence the hypercube algorithm is largely parallel. Clarifications: $^*$ This is the probability that the first round of slicing tests yields $\sigma>1$ in more than one direction. $^{**}$ This is an upper bound on the probability that positive samples may remain unidentified after the second round. Because $F^{(2)}$ is so small for $d_3=0$ and $d_3=1$, the reported value is effectively the probability that $d_3>1$. With some effort, our analysis could be extended to $d_3>1$, and to a third round of slicing tests, to further reduce the bound on the failure rate. } 
 \label{fig:tablefr}
\end{figure}

If $d_3=1$, all unknown positive sample coordinates reside in a hypercuboid with volume $3\times 2^{d_2}$. We divide this into three along the single direction with $\sigma=3$, forming three hypercuboids with $L=2$. In this case, we obtain 
\begin{equation}
 \lambda^{(j)}= { \lambda^{(1)}\over 3\times2^{j-2}}.
\label{efr4}
\end{equation}   
We are now ready to calculate the failure rate for the second round of slicing tests (assuming failure in the first round). This is the probability that $\sigma>1$ along more than one direction after the second round. Notice that, if $d_3=0$, because we subdivide the hypercube along one $\sigma=2$ direction, only if $d_2$ is greater than two in the first round, can we fail to identify all infected individuals in the second.  Using a similar argument to that used in deriving (\ref{efr1}), we obtain
\begin{equation}
 F^{(2)}|_{d_3=0}=\sum_{d_2^{(1)}=3}^D \left[1-\sum_{d=d_2^{(1)}-2}^{d_2^{(1)}-1} {\rm Pr}(d,d_2^{(1)}-1,P_2(1|\lambda^{(2)})\right]\times {\rm Pr}(d_2^{(1)},D,P_3(2|\lambda^{(1)},d_3=0))
\label{efr5}
\end{equation}   
where $P_2(1|\lambda^{(2)})$ is obtained from (\ref{psol6}) and $P_3(2|\lambda^{(1)})$ from (\ref{psol4}). If $d_3=1$, 
\begin{equation}
 F^{(2)}|_{d_3=1}=\sum_{d_2^{(1)}=2}^{D-1} \left[1-\sum_{d=d_2^{(1)}-1}^{d_2^{(1)}} {\rm Pr}(d,d_2^{(1)},P_2(1|\lambda^{(2)})\right]\times {\rm Pr}(d_2^{(1)},D-1,P_3(2|\lambda^{(1)},d_3=1)).
\label{efr5}
\end{equation} 
For practical values of $D$ and $\lambda$, both of these values are negligible (see Table \ref{fig:tablefr}) so execution beyond round two is unnecessary. Nevertheless, for the sake of completeness, we provide the failure rate for round $j>2$, namely
\begin{equation}
 F^{(j)}=\sum_{d^{(j-1)}_2=3}^{D-j+2} \left(1-\sum_{d=d_2^{(j-1)}-2}^{d_2^{(j-1)}-1} {\rm Pr}(d,d_2^{(j-1)}-1,P_2(2|\lambda^{(j)})\right)\times {\rm Pr}(d_2^{(j-1)},D-j+2,P_3(2|\lambda^{(j-1)})).
\label{efr5}
\end{equation} 

\section{Expected number of tests}
 \label{appt}

When $d_3=0$, as noted earlier, the second round of testing involves two hypercubes of dimension $d_2^{(1)} -1$. Testing them requires testing $ 4(d_2^{(1)} -1)$ slices. If needed, the third round will involve testing $ 8(d_2^{(2)} -1)$ slices and so on. Therefore the expected number of tests conducted in the $j$'th round of slicing tests, with $j>1$, is given by
\begin{equation}
t^{(j)}=2^j\sum_{d=2}^{D-j+2}(d-1) P(d|\lambda^{(j-1)}),
\label{ent1}
\end{equation} 
where
\begin{equation}
P(d|\lambda^{(1)})= {\rm Pr}(d,D,P_3(2|\lambda^{(1)},d_3=0)),
\label{ent2}
\end{equation} 
and 
\begin{equation}
P(d|\lambda^{(j)})= {\rm Pr}(d,D-j+2,P_2(2|\lambda^{(j)})).
\label{ent2a}
\end{equation} 
The expected total number of tests conducted up to and including the $j$'th round is given by
\begin{equation}
T^{(j)}=3 D +\sum_{i=2}^j t^{(i)}.
\label{ent2b}
\end{equation} 
When $d_3=1$, however, the second round of slicing tests involves 3 hypercubes and $6 d_2^{(1)}$ slices. The third round
 involves $6$ hypercubes, $12 d_2^{(2)}$ slices and so on. 
 
 The expected number of tests in the second round is:
\begin{equation}
t^{(2)}=6 \sum_{d=1}^{D-1} d\times{\rm Pr}(d,D-1,P_3(2|\lambda^{(1)},d_3=1),
\label{ent3}
\end{equation} 
and in round $j>2$ is
\begin{equation}
t^{(j)}=3\times 2^{j-1} \sum_{d=2}^{D-j+1}(d-1) {\rm Pr}(d,D-j+2,P_2(2|\lambda^{(j-1)})).
\label{ent4}
\end{equation} 
The total expected number of tests conducted up to and including the $j$'th round is given by substituting (\ref{ent3}) and (\ref{ent4}) into (\ref{ent2b}).

When $\lambda^{(1)}=0.35$ and $D=4$, in the $98.2\%$ of cases where the first round produces $d_3=0$, an average of 12.29 tests will be conducted up to and including the second round. In the remaining 1.8\% of cases, an average of 12.44 tests will be conducted up to and including the second round.

\begin{figure}
\centering
\includegraphics[width=1\textwidth]{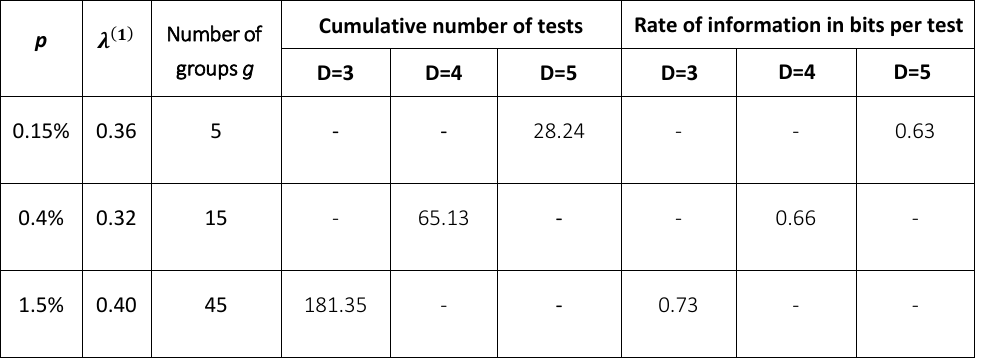}
\caption{{\bf Expected number of tests:} for $D=3,4$ and 5, in the working example presented in Section \ref{sectv}. Also shown is the ``rate" of the search, defined in Ref.~\cite{Review} as the information gained per test, in bits. Note that for the hypercube algorithm, this rate increases as $D$ falls, even exceeding that for one of the best non-adaptive algorithms known in the computer science literature (See Appendix \ref{natm} for discussion).  
 }\label{fig:tableti}
\end{figure}

Table \ref{fig:tableti} shows examples of possible field trials, as discussed in Section \ref{sectv} of the main text.  The middle three columns show the expected total number of tests needed and the right three columns show the ``rate" of the search in bits per test, as defined by Aldridge {\it et al.}, in Ref.~\cite{Review}. Note that this rate  {\it increases} with decreasing $D$. For $D=3$, it exceeds even the rate for the DD algorithm with near-constant column weight design which (for a somewhat different problem, requiring more input) achieves one of the highest rates known among known for non-adaptive test matrix algorithms (see the discussion in Appendix \ref{natm}). 

\section{Estimating the viral prevalence $p$ from test results }\label{appp}

Let $p$ be an estimate of the viral prevalence in the considered population and let $\lambda^{(0)}=p n$ be the expected number of positive samples, among $n$ samples taken from the population. Before applying the hypercube algorithm, we divide the $n$ samples into $g$ groups, where $g$ is chosen so that the expected number of positive samples in each group, $\lambda^{(1)}=\lambda^{(0)}/g$ is close to the optimal value of $0.35$.  Each sample is then subdivided into a number of smaller samples, to be used in subsequent pooled tests. 

The first round of tests is used to identify positive groups. The fraction of groups that test negative, $f$, can be used to re-estimate $p$ using the maximum likelihood method. The result is $p=1-f^{1/N}$ where $N$ is the group size. If this estimate differs significantly from the original one, we may decide to use a different group size $g$ in what follows, or even to repeat this first round of group tests since it is relatively inexpensive. 

Once the group testing phase is completed, the samples from every positive group are placed in a hypercube for further slicing tests. Suppose that, in the first round of slicing tests, we observe $n_1, n_2$ and $n_3$ instances of $\sigma=1,2$ and $3$, respectively, across all positive groups. The posterior probability of $k$, conditioned on these observations is then given from Bayes' theorem by
\begin{equation}
P(k|{\rm obs})= {P({\rm obs}|k) P(k)\over P({\rm obs})}\propto P({\rm obs}|k) P(k),
\label{evp1}
\end{equation}   	
where 		
\begin{equation}
P({\rm obs}|k)= \binom{n_1+n_2+n_3}{n_1,n_2,n_3} {3^{(n_1+n_1+n_3)(1-k)} } (2^k-2)^{n_2}(3^{k-1}-2^k+1)^{n_3},
\label{evp2}
\end{equation}   	
and the prior probability is given by the Poisson distribution, 
\begin{equation}
P(k)=e^{-\lambda} {\lambda^k\over k!}.
\label{evp3}
\end{equation}   	
The normalisation factor $P({\rm obs})$ is then determined by requiring that $\sum_{k=0}^\infty P(k|{\rm obs})=1$. The required summations converge rapidly at large $k$. Using (\ref{evp1}) the expected value of $k$ is determined,
\begin{equation}
\overline{k} = \sum_{k=0}^\infty P(k|{\rm obs}) k,
\label{evp4}
\end{equation}   	
and the estimate of the prevalence is $p=\overline{k}/n$ where $n$ is the initial population sample size. A credible interval ({\it e.g.}, a 95\% Cl) for $k$ and hence for $p$ can be easily computed from (\ref{evp1}). A refinement, taking into account that the triples are taken from a set of $g$ indistinguishable groups of size $N$, will be presented elsewhere.

\section{Experimental Methods and Supplementary Information}\label{methods}

\noindent {\bf Observational study design:} 
We conducted an experiment to evaluate the hypothesis that known SARS-CoV-2 positive oropharyngeal swab specimens collected during COVID-19 surveillance in Rwanda will test positive after they are combined with as many as 99 known SARS-CoV-2 negative specimens. This was followed by an observational study that aimed to apply our hypercube algorithm to increase the efficiency of community testing for COVID-19 in Rwanda. In the experiment, two different sets of sample pools were tested for SARS-CoV-2 using RT-PCR. Each set consisted of three sample pools containing one known SARS-CoV-2 positive sample diluted in ratios of 1:20, 1:50, and 1:100 by combining it with equivalent amounts of 19, 49, and 99 known SARS-CoV-2 negative samples, respectively (see Fig.~\ref{fig2}). In the observational study, 1280 individuals selected from the community were tested for SARS-CoV-2 using RT-PCR. One third of the individuals were participants in a screening for Severe Acute Respiratory Infections (SARI) and Influenza Like Illness (ILI) conducted in 30 per cent of the health facilities found across the 30 districts of Rwanda. The remaining two thirds were from COVID-19 screening of at-risk groups in the capital city of Kigali. The latter group is comprised mainly of people (market vendors, bank agents, and supermarket agents) who remained active during the lockdown imposed by the Government of Rwanda to contain COVID-19. Figure   \ref{fig:table} summarises the characteristics of the study participants.

\begin{figure}
\centering
\includegraphics[width=1\textwidth]{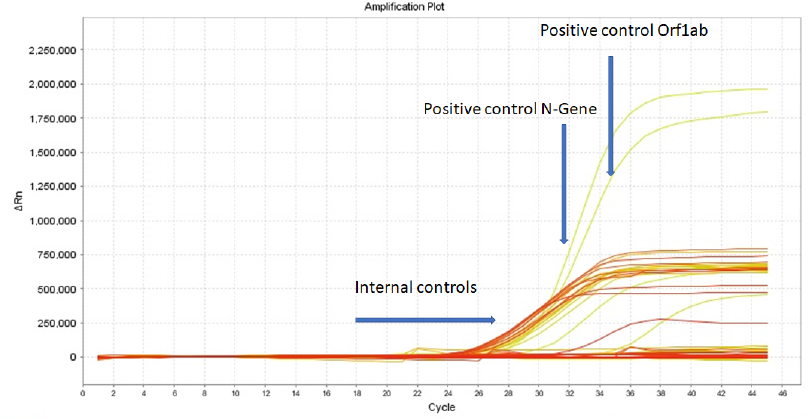}
\caption{{\bf Amplification plot for sample pools.}
Each of 64 the sample pools described in the text tests negative for SARS-CoV-2: the RT-PCR fluorescence curves show below-threshold net fluorescence values. In contrast, for both target genes of the positive control, the fluorescence curves cross the threshold after 32 PCR cycles. $\Delta Rn$ denotes the difference between the fluorescence signal generated by a sample and a baseline signal.  The yellow curves reaching $\Delta R n \sim 2,000,000)$ and $1,750,000$ represent the positive control for the N and Orf genes, respectively. The other yellow and orange curves represent internal controls.  }\label{fig:suppfig}
\end{figure}

\begin{figure}
\centering
\includegraphics[width=1\textwidth]{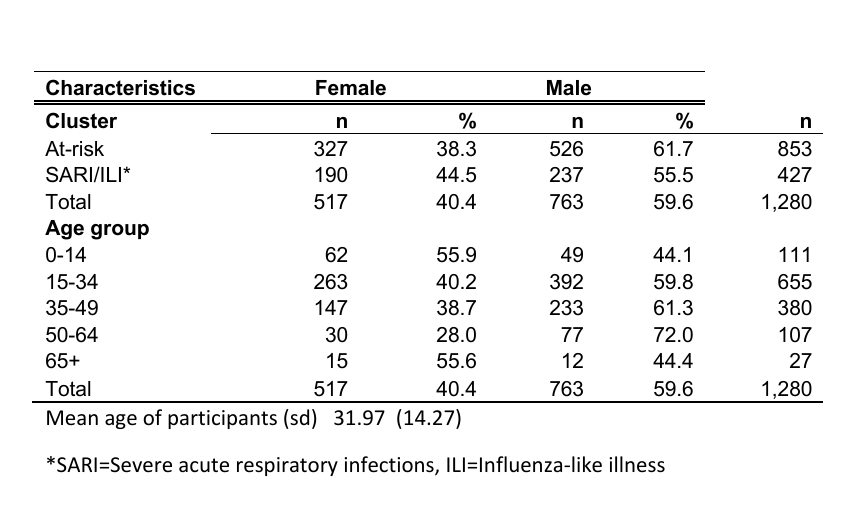}
\caption{{\bf Characteristics of participants in field trial of hypercube algorithm in Rwanda.} \hfill\eject For more information, see Observational study design. }\label{fig:table}
\end{figure}

The positive fraction of RT-PCR tests for SARS-CoV-2 conducted in Rwanda in March 2020 suggests an upper-bound of 2 per cent for the virus prevalence in the country. Using $p=$2 per cent in the hypercube algorithm indicated an optimal sample group size of 17.5. For convenience, the 1280 individual samples were combined in 64 groups of 20 samples before testing for SARS-CoV-2 (see Fig. \ref{fig:suppfig} for the experimental results).

We used two established experimental protocols for SARS-CoV-2 testing, namely 1) a protocol by DAAN Gene Co., Ltd., Sun Yat-sen University, which is available online~\cite{Daan}, and is also under review by the WHO~\cite{WHO}, and 2) another by Corman {\it et al.},~\cite{Corman2020} which is widely used by the scientific community. The first protocol is used for routine screening for SARS-CoV-2, while the second protocol is used only if the first one produces a positive result and confirmation is thus required.

\noindent {\bf Sample collection and pool design:}  Oropharyngeal swabs were collected by wiping the tonsils and posterior pharynx wall with two swabs, and the swab heads were immersed in 3 ml Viral Transport Medium (VTM). Samples were transported in VTM to the Rwanda National Reference Laboratory (NRL) immediately after collection. Samples that had to be transported over a long distance were stored in dry ice. Each sample had a volume of 3 ml, of which 200 $\mu l$ were used for pool testing, and the remainder was temporarily stored at -20°C until the result of the pool testing was known. 200 $\mu l$ of each sample were mixed with the same volume of other samples of the same pool in a FalconTM 15 ml conical tube and, after vortexing for 5 seconds, 200 $\mu l$ of the  mixture were pipetted for downstream RNA extraction. 5 $\mu l$ of the extracted RNA were added to 20 $\mu l$ of master mix to make 25 $\mu l$ of total solution to be amplified by RT-PCR. If a pool tested positive, stored samples from that pool were processed to identify the positive ones. Individual samples were bar coded, making it easy to trace individuals that tested positive and minimising the risk of confusion of samples. Pool design and subsequent experimental analysis (see {\bf RT-PCR for SARS-CoV-2} below) were implemented with the aid of a robot to reduce human error.


\noindent {\bf RT-PCR for SARS-CoV-2:}  Total viral RNA was extracted from swab specimens using the QIAamp Viral RNA 91 Mini Kit (Qiagen, Hilden, Germany), according to the manufacturer’s instructions. RNA samples were screened for SARS-CoV-2 using a 2019-nCoV RNA RT-PCR test targeting two genes respectively encoding an open reading frame (denoted Orf1ab) and nucleocapsid protein (denoted N) (DAAN Gene Co., Ltd. Of Sun Yat-sen University, 19, Xiangshan Road, Guangzhou Hi-Tech Industrial Development Zone, China). For Orf1ab, CCCTGTGGGTTTTACACTTAA and ACGATTGTGCATCAGCTGA were used as forward and reverse primers, respectively, together with a 5'-VIC CCGTCTGCGGTATGTGGAAAGGTTATGG-BHQ1-3' probe. For N, GGGGAACTTCTCCTGCTAGAAT and CAGACATTTTGCTCTCAAGCTG were used as forward and reverse primers, respectively, together with a 5'-FAM- TTGCTGCTGCTTGACAGATT-TAMRA-3' probe. The RT-PCR reaction was set up according to the manufacturer’s protocol, with a total volume of 25 $\mu$L. The reaction was run on the ABI Prism 7500 SDS Instrument (Applied Biosystems) at 50°C for 15 min for reverse transcription, denatured at 95°C for 15 min, followed by 45 PCR cycles of 94°C for 15 sec and 55°C for 45 sec. A threshold cycle (Ct value) $<$40 indicated a positive test, while Ct value $>$40 indicated a negative test. Positive controls for the reaction showed amplification as determined by curves for FAM and VIC detection channels, and a Ct value $\leq$ 32. Positive tests were confirmed using LightMix SarbecoV E-gene and LightMix Modular SARS-CoV-2 RdRp RT-PCRs targeting the envelope (E) and RNA directed RNA Polymerase (RdRp) genes, respectively, as described by the manufacturer (TIB MOLBIOL Syntheselabor GmbH, Eresburgstr. 22-23, D-12103 Berlin, Germany). Both the primers used and the RT-PCR reaction conditions were previously described~\cite{Corman2020}.

\noindent {\bf Statistical analysis:} Ct values were tested for normality by using the Shapiro-Wilk test. A confidence bound for a sample of $n$ Ct values was calculated as ${\bar{C_t}} \pm {t}^{*}_{df} \times s $, where $\bar{C_t}$ is the sample mean, $s$ is the sample standard error, and ${t}^{*}_{df}$ is an appropriate quantile of the Student’s $t$ distribution with $n-1$ degrees of freedom, $df$. A confidence bound for the sum of the means of two samples of Ct values of sizes $n_1$ and $n_2$, respectively, was calculated using the same formula, with $\bar{C_t}$ set to the sum of the individual sample means, $s$ set to the sum of the individual sample standard errors, and $df$ set to the smaller of $n1-1$ and $n2-1$. Statistical analysis was done using the R statistical computing environment (https://www.r-project.org/).

\noindent {\bf Ethics approval:} 
Ethics approval was obtained from the Rwanda National Ethics Committee (Ref: FWA Assurance No. 00001973 IRB 00001497 of IORG0001100/20March2020) and written informed consents were obtained from the patients.

\noindent {\bf Data availability:}  All data are available from the corresponding authors upon reasonable request.

\begin{figure}
\centering
\includegraphics[width=1\textwidth]{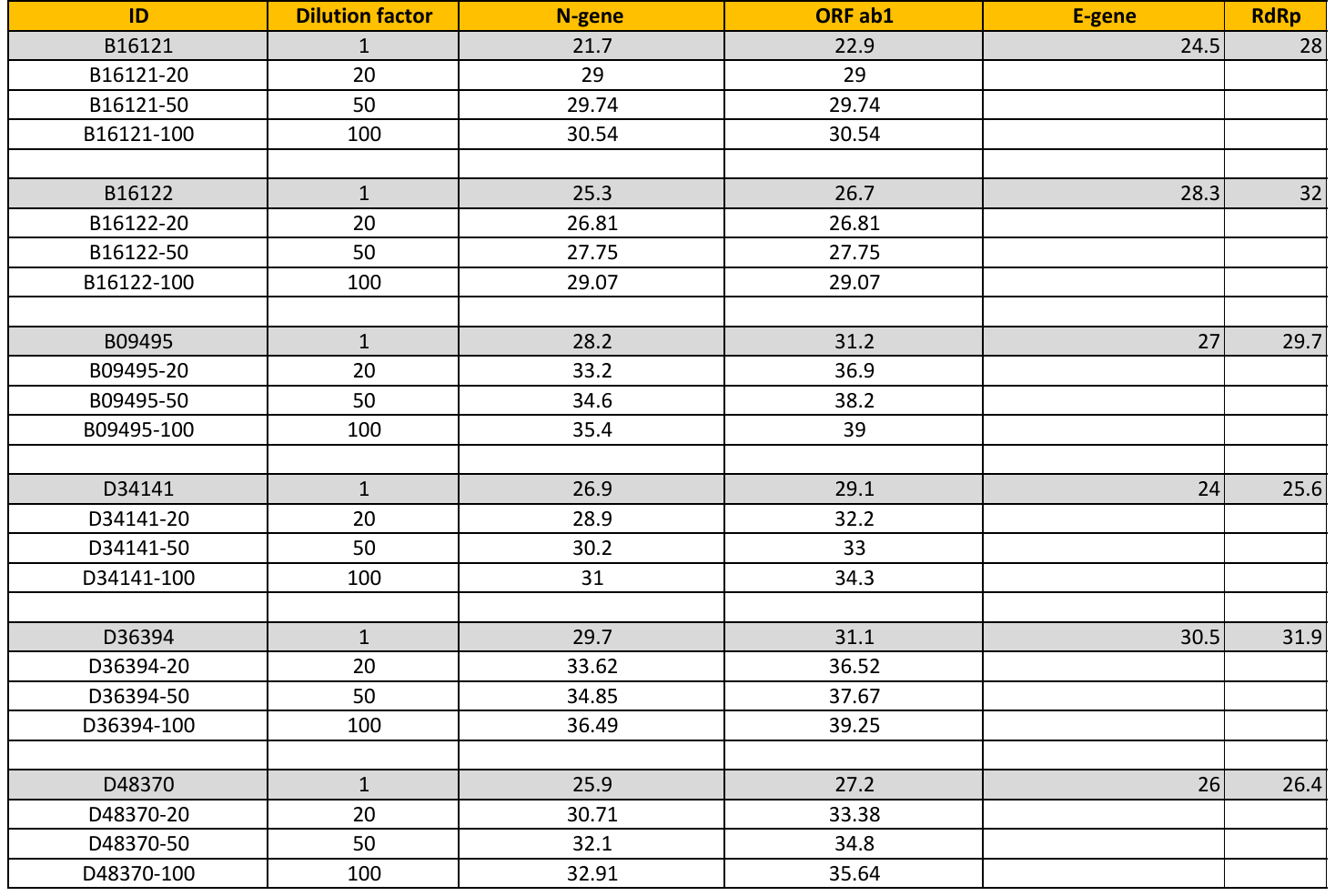}
\caption{{\bf Positive specimens are detected after massive dilution.} Six SARS-CoV-2 positive specimens detected during COVID-19 screening in Rwanda were analysed. The positive specimens were detected by using a screening RT-PCR test targeting the N and Orf1ab genes of SARS-CoV-2 (Ct values from this test are reported in columns 3 and 4), and confirmed by using another RT-PCR test targeting the E and RdRp genes (Ct values reported in columns 5 and 6). We determined whether the screening test would have detected the positive specimens if they had been combined with 19, 49 or 99 known SARS-CoV-2 negative specimens. Three pools were thus formed per sample, with dilution factors given in column 2. For all 18 pools, fluorescence from RT-PCR reactions exceeded background levels at Ct values below 40 (columns 3 and 4), implying that all the positive samples would have been detected even if they were diluted by up to 100 fold. See Methods for details of the experimental methods.}\label{fig:Stable1}
\end{figure}

\begin{figure}
\centering
\includegraphics[width=1\textwidth]{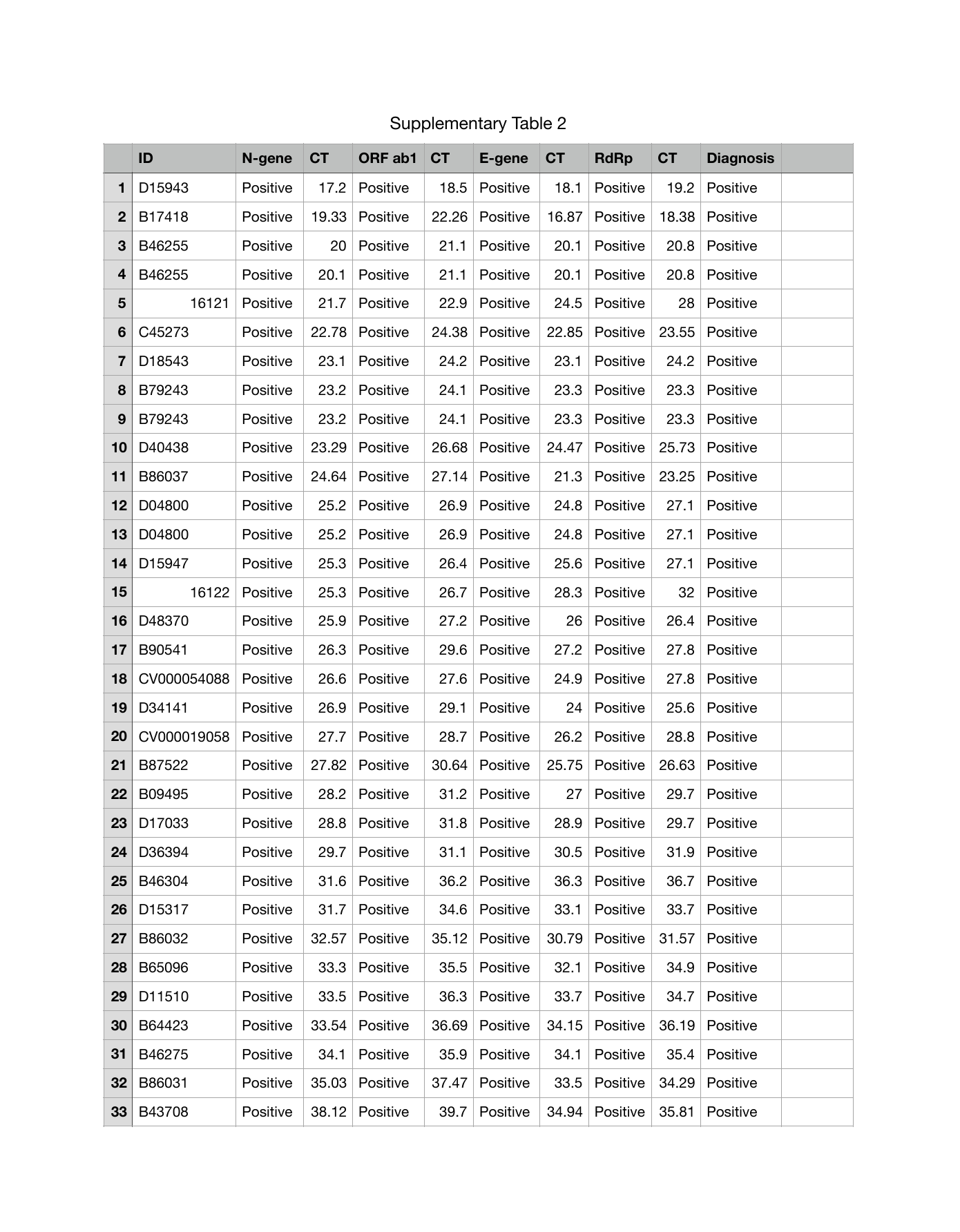}
\caption{{\bf Representative sample of confirmed positive specimens detected during COVID-19 screening in Rwanda.} The specimens were detected by using a screening RT-PCR test targeting the N and Orf1ab genes of SARS-CoV-2 (Ct values from this test are reported in columns 3 and 5). They were subsequently confirmed as positive by using another RT-PCR test targeting the E and RdRp genes (Ct values are reported in columns 7 and 9).}\label{fig:Stable2}
\end{figure}

\end{document}